\documentclass[aip,preprint]{revtex4-1}

\usepackage{graphicx}
\usepackage{verbatim}

\draft 

\begin{document}

\preprint{}

\title{Dead leaves and the dirty ground: low-level image statistics in transmissive and occlusive imaging environments}

\author{Joel Zylberberg}
\affiliation{Department of Applied Mathematics, University of Washington, Seattle WA 98195}
\affiliation{Department of Physics, University of California, Berkeley CA 94720}
\affiliation{Redwood Center for Theoretical Neuroscience, University of California, Berkeley CA 94720}

\author{David Pfau}
\affiliation{Redwood Center for Theoretical Neuroscience, University of California, Berkeley CA 94720}
\affiliation{Center for Theoretical Neuroscience, Columbia University, New York NY 10032}

\author{Michael Robert DeWeese}
\email[]{deweese@berkeley.edu}
\affiliation{Department of Physics, University of California, Berkeley CA 94720}
\affiliation{Redwood Center for Theoretical Neuroscience, University of California, Berkeley CA 94720}
\affiliation{Helen Wills Neuroscience Institute, University of California, Berkeley CA 94720}

\begin{abstract}
The opacity of typical objects in the world results in occlusion --- an important property of natural scenes that makes inference of the full 3-dimensional structure of the world challenging. The relationship between occlusion and low-level image statistics has been hotly debated in the literature, and extensive simulations have been used to determine whether occlusion is responsible for the ubiquitously observed power-law power spectra of natural images. To deepen our understanding of this problem, we have analytically computed the 2- and 4-point functions of a generalized ``dead leaves" model of natural images with parameterized object transparency. Surprisingly, transparency alters these functions only by a multiplicative constant, so long as object diameters follow a power law distribution. For other object size distributions, transparency more substantially affects the low-level image statistics. We propose that the universality of power law power spectra for both natural scenes and radiological medical images -- formed by the transmission of x-rays through partially transparent tissue --
stems from power law object size distributions, independent of object opacity.
\end{abstract}

\pacs{42.30.Va, 89.75.Da, 89.75.Kd, 87.57.-s,  42.66.Lc, 05.70.Jk}

\date{\today}

\maketitle

\section{Introduction}

Natural images are surprisingly statistically uniform. The autocorrelation function, a measure of how similar nearby pixels tend to be, is virtually universal for natural images~\cite{stephens,ruderman,dong_atickB,field,olshausen_simoncelli,torralba,vandersschaff} (Fig.~1). This is typically quantified by measuring image power spectra (Fourier transform of the autocorrelation function),
which are well-described by scale-invariant power law functions with power $\mathcal{P}$ and spatial frequency $k$ related by $\mathcal{P}(k)\propto k^{-\alpha}$, with exponents $\alpha \approx 2$. The exponents $\alpha$ vary slightly from image-to-image, and there are small differences in average exponent $\alpha$ between terrestrial~\cite{field,ruderman,vandersschaff} and aquatic~\cite{balboa} environments, and between natural and man-made ones~\cite{torralba}. 

Intriguingly, even radiological images like mammograms have power law power spectra~\cite{heine_velthuizen,li}, typically with larger $\alpha$ values, despite the fact that the physics of image formation are very different for radiological and natural images. In natural images, formed by reflection of light off of surfaces, objects tend to be opaque, and thus they occlude one another, whereas in mammograms, formed by the transmission of x-rays through breast tissue, objects are more transmissive and do not completely occlude one another.
The statistics of radiological images have received less attention and are less well understood. Interestingly, however, the powers $\alpha$ typically vary between mammogram images of patients with low vs. high risk of developing breast cancer~\cite{li}, and vary as a function of the density of the breast tissue~\cite{metheany}, highlighting the potential clinical importance of these image statistics.

The statistical regularity of natural scenes implies that engineers can design, and evolution might have selected for, coding schemes that exploit this structure~\cite{barlow61,olshausen_simoncelli}. 
Indeed, the peripheral mammalian visual system appears to exploit this homogeneity by using simple filters to decorrelate the incoming signal~\cite{dong_atickA,atick_redlich,dan} and more complex feature dictionaries to efficiently encode the decorrelated signal~\cite{jz_plos,olshausen_simoncelli,rehn_sommer}.
\begin{figure}[b!]
\begin{center}
\includegraphics[width=3.6in]{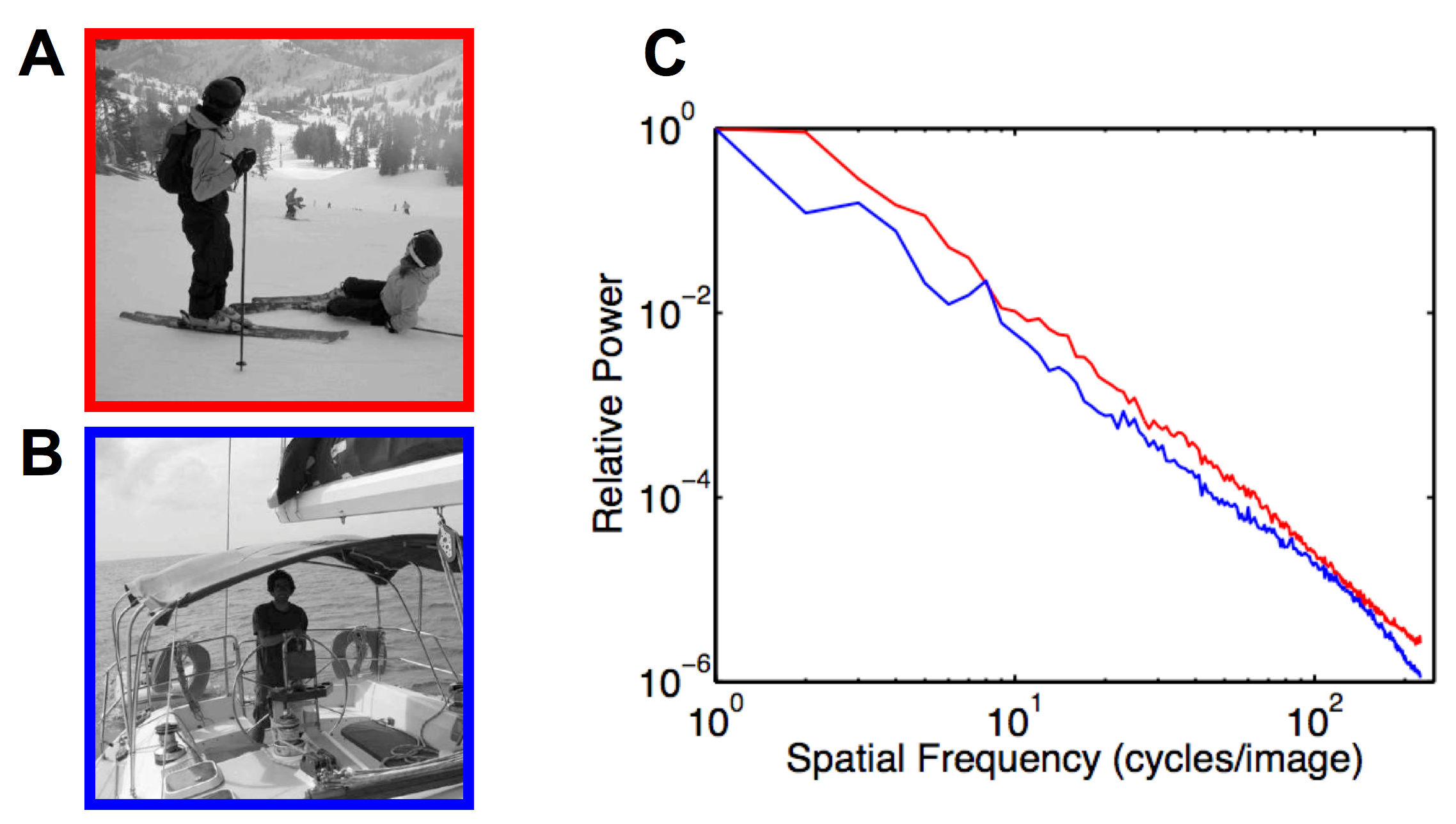}
\caption{ (Color online)  \textbf{Natural images have nearly identical scale invariant power spectra.} Even very different natural images (\textbf{A},\textbf{B}) have similar rotation-averaged power spectra that each follow a power law (\textbf{C}). Line colors in panel \textbf{C} match the borders of corresponding panels \textbf{A} and \textbf{B}. The upper curve in panel \textbf{C} corresponds to the image in panel \textbf{A} while the lower one corresponds to \textbf{B}. }
\end{center}
\end{figure}

Using the intuition that the environment is composed of distinct objects, Ruderman studied a ``dead leaves" model~\cite{matherton_1968,bordenave} for natural scenes, in which images are created by sequentially placing opaque, potentially overlapping circles of random brightness in random locations on a 2-dimensional image plane~\cite{ruderman_scaling} (Fig.~2A). Ruderman modeled correlations between pixels by assuming a different correlation function for points falling within a visible circle than for points falling in different visible circles. Using analytical calculations he demonstrated that, so long as the diameters $s$ of the circles follow a power law distribution with probabilities $p(s) \propto s^{-(3 + \eta)}$,
the images exhibit power law correlation functions, $C(q) \propto q^{-\eta}$, where $q$ is the separation between pixels, and power law power spectra, $\mathcal{P}(k) \propto k^{-(2-\eta)}$.
If the circle sizes are drawn from other distributions, Ruderman's analytical calculations suggest that the power spectra that could be made to differ from a power law, contrary to the old notion~\cite{carlson} that the $1/k^2$ power spectra result from the mere presence of edges, each of which has a $1/k^2$ 1-dimensional power spectrum (\emph{cf.} Balboa et al.~\cite{balboa_2001}). More recently, Balboa et al.~\cite{balboa_2001} simulated the analytical examples presented by Ruderman~\cite{ruderman_scaling},
including images with the exponential distribution of object sizes that was claimed~\cite{ruderman_scaling} to yield non-power-law power spectra. They found that these images had nearly power law power spectra, and subsequently reiterated the previous claim that occlusion, and not object size distributions, are the cause of power law power spectra in natural images.

This ``edges vs. size distributions" debate was subsequently resolved when Hsiao and Milane demonstrated, via numerical simulations, that dead leaf models with partially transparent objects (and thus only partial occlusion) whose sizes follow a power law distribution yield power law power spectra, and that dead leaf models with opaque objects from other size distributions can have non power-law power spectra~\cite{hsiao}. In other words, occlusion is neither necessary, nor sufficient, to yield power law image power spectra. In the same paper, Hsiao and Milane computed the power spectrum of a simplified ensemble of images formed by summing the intensities of different randomly placed disks. This model was simpler than the images with partially occluding leaves that they simulated. The linearity of this model makes it relatively straightforward to compute the Fourier transform of the model images, and thus to estimate the power spectra.

Thus, to date, the 2-point statistics of dead leaf image models have been analytically calculated for both fully opaque leaves~\cite{ruderman_scaling}, 
and for fully transmissive leaves~\cite{hsiao}.
What remains is to solve for the 2-point function of images with partial occlusion,  which will deepen our understanding of how opacity and image statistics inter-relate along this continuum of object properties. Thusly motivated, we studied a generalized dead leaves model, in which the leaves have variable transparency. While general feature probabilities have been solved exactly for the fully opaque dead leaves model~\cite{pitkow}, our transparent generalization requires other methods and has not previously been systematically explored. We show herein that, so long as leaf sizes follow a power-law distribution, transparency results in an overall multiplicative factor in the 2- and 4-point functions but does not change their functional (power-law) form. For other size distributions, transparency does change the form of the autocorrelation function, suggesting that power-law size distributions, unify the observed power spectra of natural and radiological images. 

\section{Analytical calculation of the 2-point function in the transmissive dead leaves model}


We begin by analytically computing the 2-point functions of images in our ``transmissive dead leaves" environment.  For image pixels values $I(\vec{x})$, the 2-point function is given by $C(\vec{x},\vec{x}') = \left< I(\vec{x})  I(\vec{x}') \right> = C(|\vec{x}-\vec{x}'|)$, where the angle brackets denote averaging over images drawn from this ensemble 
and the second step stems from the fact that, since our model world is invariant under both translations and rotations, the 2-point function depends only on the distance $|\vec{x}-\vec{x}'| = q$ between sample points. 

The image is formed by randomly placing a circle whose diameter $s$ is drawn from some distribution, with brightness value $b$, and transparency $a$, on a surface of diameter $L$. The brightnesses $b$ will be drawn from a zero-mean distribution, and the transparencies $a \in [0,1]$ can also be random. A value $a=1$ specifies a fully transparent (invisible) circle, while a value of $a=0$ specifies a fully opaque circle, as in Ruderman's model~\cite{ruderman_scaling}. When a new circle is added, the pixel value $I(\vec{x})$ at a point $\vec{x}$ that falls within the circle undergoes the transformation
\begin{equation}
I(\vec{x}) \to (1-a)b + aI(\vec{x}).
\end{equation}
Pixels not lying under the circle are unaffected by its addition. This process is continued ad infinitum to create model images (Fig.~2).
 \begin{figure}[ht!]
\begin{center}
\includegraphics[width=3.6in]{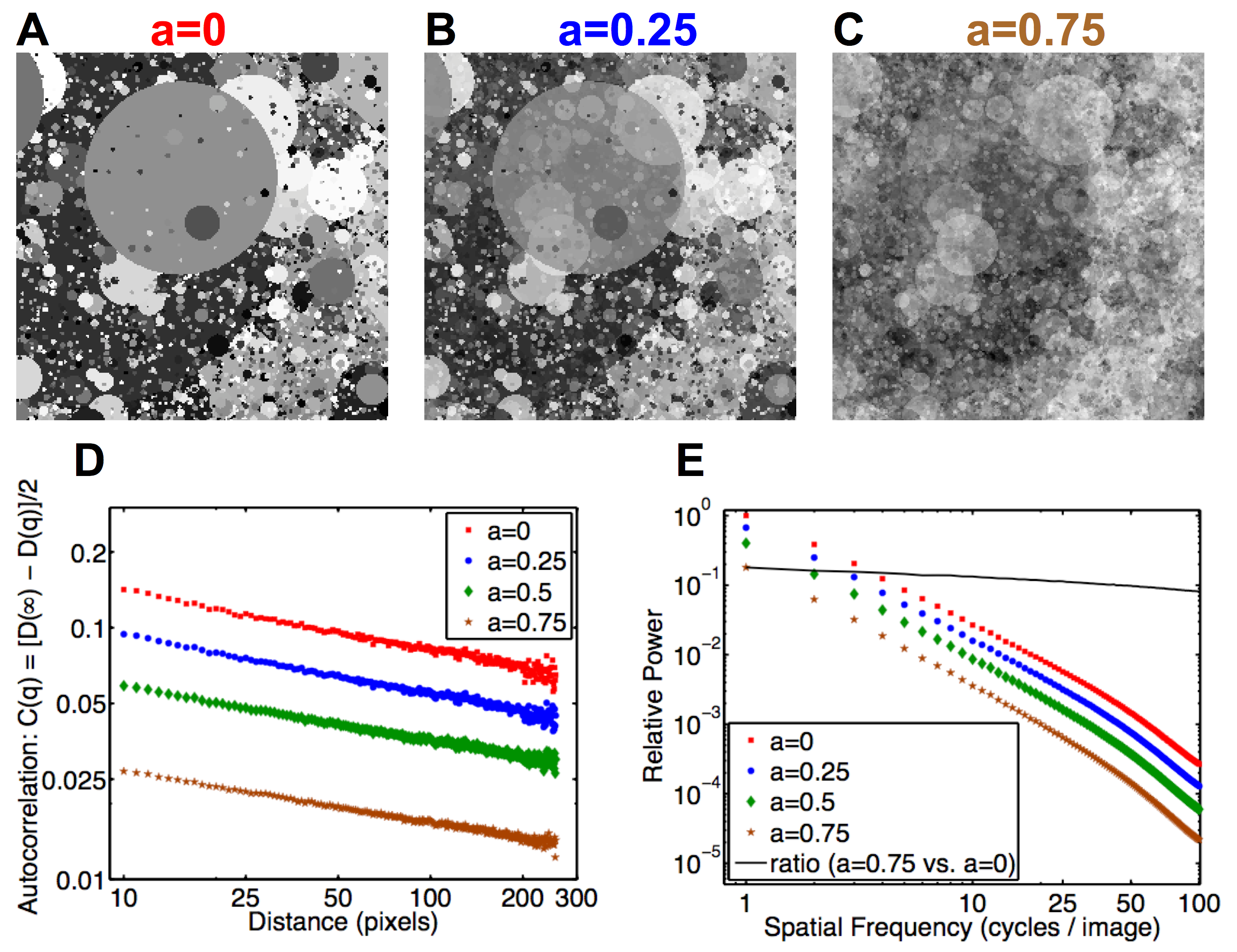}
\caption{ (Color online) \textbf{For power law object size distributions, the 2-point statistics of opaque and transmissive dead leaves images differ by a multiplicative constant.} (\textbf{A}) A representative image from the opaque ($a=0$) dead leaves model
with circle diameters drawn from the distribution $p(s) \propto s^{-3.2}$ for $s>s_0 = 1$ pixel and circle brightnesses drawn uniformly within $b \in [-1,1]$. (\textbf{B}) When the circles are partially transparent ($a=0.25$ for all circles), but all other parameters are the same, previously occluded circles are partially visible. (\textbf{C}) A higher level of transparency ($a = 0.75$) results in an image that begins to approximate Gaussian pink noise, as expected from the central limit theorem~\cite{pitkow}. (\textbf{D}) Autocorrelation functions of dead leaves image ensembles of different opacity levels differ only by a multiplicative constant for power-law object size distributions. 
The 2-point functions are power law functions of distance, with power $\sim-0.2$, in good agreement with our analytical calculation.
(\textbf{E}) Similarly, the power spectra of these image ensembles are roughly power-law functions
and are all the same up to a multiplicative constant. The ratio of the opaque and most transparent power spectra is nearly flat. At relatively high spatial frequencies (above $\sim 20$ cycles/image), corresponding to small length scales, the $q \gg s_0$ approximation in our analytical calculation fails, and slight deviations from power-law power spectra can be observed, as can deviations from constancy in the ratio.}
\end{center}
\end{figure}

We will compute $\left< I(\vec{x})^2 \right>$ and $C(q)$ recursively by noting that adding another leaf to an image creates a new image from the same transmissive dead leaves ensemble and thus the (average) statistical properties must remain unchanged by this transformation~\cite{ruderman_scaling}. 

Using Eq.~(1), we can compute the pixel variance
\begin{eqnarray}
\left< I^2(\vec{x}) \right> &=& \left( 1-P_{in} \right) \left< I^2(\vec{x}) \right>   \\
&+& P_{in} \left< \left( aI(\vec{x})  + (1-a)b \right)^2   \right>   \nonumber \\ \nonumber
\Rightarrow \left< I^2(\vec{x}) \right> &=& \frac{ \left< b^2 \right>  \left< (1-a)^2\right>} {1- \left< a^2 \right> } \nonumber,
\end{eqnarray}
where $P_{in}$ is the probability that the point in question falls within the newly added circle. The quantity $P_{in}$, and thus the distribution of circle sizes, does not affect the pixel variance. It will however, affect the spatial properties of the image, including $C(q)$. 

To compute $C(q)$, consider how the pixel values of a pair of points with separation $q$ are affected by the addition of a new leaf. After adding the leaf, either one, both, or neither of the sample points lie under the leaf, resulting in three different possible modifications to the pixel values (Eq.~1). These outcomes occur with probabilities $P_1(q)$, $P_2(q)$, or $P_0(q)$, respectively, which we will later compute. Equating the 2-point functions before and after the addition of a new leaf, we obtain
\begin{eqnarray}
&C(q)& = P_0(q) C(q) + P_1(q)  \left<  \left[ a I(\vec{x}) + (1-a)b    \right]  I(\vec{x}') \right>  \nonumber \\
					&+& P_2 (q) \left<  \left[ a I(\vec{x}) + (1-a)b    \right]  \left[ a I(\vec{x}') + (1-a)b    \right]  \right>. 
\end{eqnarray}
Recalling the definition of the autocorrelation function and the normalization $P_0 (q)+ P_1(q) + P_2(q) = 1$, we find
\begin{equation}
 C(q) =   \frac{  \left< b^2 \right> \left< (1-a)^2 \right> P_2(q)}{P_1(q)\left<  1-a  \right> + P_2(q)   \left< 1-a^2  \right> }. 
\end{equation}
The quantities $\left< b^2 \right>$,  $\left< a^2 \right>$, and $\left< a \right>$ depend on the distributions of circle brightnesses and opacities.

To calculate $P_1(q)$, we first define $P^{\star} = \left< s^2 \right> / L^2$, which is the probability that any given point in the image falls within a newly-deposited leaf. Here $L$ is the diameter of the circular image area, $s$ is the diameter of the newly added circle, and we assume $\left<s^2 \right> \ll L^2$. The probability $P_1(q)$ that either point, but not both, falls within the circle is then $P_1(q) = 2 \left( P^{\star} - P_2(q) \right)$, where the factor of 2 comes in because there are two such points to consider.

To determine the probability $P_2(q)$, note that, for a circle of diameter $s$, given that one particular point $\vec{x}$ is within the circle (which occurs with probability $s^2/L^2$), the probability that another point, a distance $q$ away, is also within the circle, is given by~\cite{ruderman_scaling} $g(q/s \in [0,1]) = \frac{2}{\pi} \left[ \cos^{-1}(q/s) - (q/s) \sqrt{1-(q/s)^2} \right]$, and thus
\begin{equation}
P_2(q) = \int_{0}^{\infty} \frac{s^2}{L^2} g(q/s)p(s) ds.
\end{equation}
For a power law size distribution $p(s) = (A/s_0) (s/s_0)^{-\alpha}$, where $\alpha > 3$, $A$ is a unitless normalization constant, and $s_0$ is the small-size cutoff, the change of variables $u = s/q$ in the above integral yields
\begin{equation}
P_2(q) =  A \left( \frac{s_0}{L} \right)^2 \left( \frac{q}{s_0} \right)^{-(\alpha -3)}  \int_{1}^{\infty}  g(1/u) u^{2-\alpha} du.
\end{equation}
Define the integral to be $B(\alpha)$. For pixel separations much larger than the small-size cutoff of our leaf diameter distribution, $q \gg s_0$ (in which case $P^{\star} = \frac{A}{\alpha-3} \left( \frac{s_0}{L}  \right)^2 \gg P_2(q)$), Eq.~(4) becomes
\begin{equation}
C(q) = \frac{   B(\alpha) \left(\alpha-3 \right) \left< b^2 \right> \left< (1-a)^2  \right>}{2\left<  1-a \right>}  \left( \frac{q}{s_0} \right)^{-(\alpha - 3)}, 
\end{equation}
yielding an image power spectrum~\cite{ruderman_scaling} $\mathcal{P}(k) \propto \frac{ \left< b^2 \right> \left< (1-a)^2  \right>}{\left<  1-a \right>} k^{-(5-\alpha)}$ in which the opacity affects the power spectrum only as a multiplicative prefactor. When $a=0$ for all circles (opaque limit), our result is equal to that of Ruderman~\cite{ruderman_scaling}, as it must be. Also note that, as one might expect, the 2-point function does not depend on the size $L$ of the image surface.

To demonstrate that leaf opacity can affect the functional form of the 2-point function, we repeat the above calculations, but now have all leaves be the same size $s^{\star}$. The size distribution is thus $p(s) = \delta(s-s^{\star})$, in which case the correlation function is
\begin{equation}
C_{\delta} (q) =  \frac{ \left<b^2 \right> \left<\left( 1-a \right)^2 \right>  g(q/s^{\star})}{2 \left<1-a \right> -  \left< \left(1+a\right)^2 \right> g(q/s^{\star})},
\end{equation}
which depends non-trivially on $a$: for $q>s^{\star}, g(q/s^{\star})=0$ and the correlation function vanishes, so the large-$q$ limit in which Eq.~(7) was derived is irrelevant for delta-function size distributions. Furthermore, even for fully opaque leaves, it is clear that this correlation function, which is identically zero for  $q>s^{\star}$, is not described by a power-law function of distance. 

A comparison of Eqs. (2) and (7) shows that the pixel variance, and the image autocorrelation function, are multiplied by different opacity dependent pre-factors. For $q=0$, the variance and the 2-point function are equal, so the fact that for $q \gg s_0$, they scale differently with changing opacity highlights that there is a qualitative change in the 2-point function near the $q \sim s_0$ boundary. For natural images, the minimum object size is much smaller than our cameras can resolve, so this boundary is never encountered in practice. Furthermore, this comparison demonstrates that not all image statistics vary  in the same way with changing leaf opacity.

\section{Analytical calculation of the 4-point function for collinear points in the transmissive dead leaves model}

As we have seen, the form of the 2-point function is independent of leaf opacity for power-law object size distributions. At the same time, the images generated with different leaf opacities (Fig. 2) are visibly different, so there must be some difference in the image statistics (aside from the overall pixel variance) from ensembles with different object opacities. To understand this difference, we consider higher-order statistics beyond the 2-point function. If the leaf brigthnesses $\left<b \right>$ are symmetrically distributed about zero, then the 3-point function will vanish, and so the next possible candidate beyond the 2-point function is the 4-point function.

In this section, we will compute the 4-point function $C^{coll}_4(\vec{x}, \vec{x}',\vec{x}'',\vec{x}''') = \left<  I(\vec{x}) I(\vec{x}') I(\vec{x}'') I(\vec{x}''')  \right>$ for equidistant collinear points; $|\vec{x}-\vec{x}'| = |\vec{x}'-\vec{x}''| = |\vec{x}''-\vec{x}'''| = q$ and $|\vec{x}-\vec{x}''| = |\vec{x}'-\vec{x}'''| = 2q$, for the dead leaves model with power-law leaf size distribution. We chose this arrangement of points because it considerably simplifies the analysis of the 4-point function, for reasons that will become apparent during the calculation. Nevertheless, the calculation itself is still somewhat tedious, so some readers may wish to skip to the result at the end of this section. 

As in the case of the 2-point function described above, since our image ensemble is invariant under translations and rotations, the result depends only on the pixel spacing $q$: $C^{coll}_4(\vec{x}, \vec{x}',\vec{x}'',\vec{x}''')  = C^{coll}_4(q)$. We apply the same recursive logic that we used for computing the 2-point function in order to infer the 4-point function, and start by enumerating all of the possible modifications to the 4-point function upon the addition of a new circle. We will number the points from left to right. 
Thus,

\begin{eqnarray}
C^{coll}_4(q) &=& P^{coll}_{\o}(q) C^{coll}_4(q)   \\
&+& P^{coll}_{1,\o} (q)  \left<  \left[ a I(\vec{x}) + (1-a)b    \right]  I(\vec{x}') I(\vec{x}'') I(\vec{x}''')  \right>  \nonumber \\
&+& P^{coll}_{2,\o}(q)  \left<  I(\vec{x}) \left[a I(\vec{x}') + (1-a)b    \right] I(\vec{x}'') I(\vec{x}''')  \right> \nonumber \\
&+& P^{coll}_{3,\o}(q)  \left<  I(\vec{x}) I(\vec{x}')  \left[a I(\vec{x}'') + (1-a)b    \right] I(\vec{x}''')  \right> \nonumber \\
&+& P^{coll}_{4,\o}(q)  \left<  I(\vec{x}) I(\vec{x}')  I(\vec{x}'')  \left[a I(\vec{x}''') + (1-a)b    \right]  \right> \nonumber \\
&+& P^{coll}_{1,2}(q)  \left<  \left[a I(\vec{x}) + (1-a)b    \right]\left[a I(\vec{x}') + (1-a)b    \right] I(\vec{x}'')  I(\vec{x}''')  \right> \nonumber \\
&+& P^{coll}_{2,3}(q)  \left< I(\vec{x})  \left[a I(\vec{x'}) + (1-a)b    \right]\left[a I(\vec{x}'') + (1-a)b    \right]  I(\vec{x}''')  \right> \nonumber \\
&+& P^{coll}_{3,4}(q)  \left< I(\vec{x})  I(\vec{x}')  \left[a I(\vec{x''}) + (1-a)b    \right]\left[a I(\vec{x}''') + (1-a)b    \right]   \right> \nonumber \\
&+& P^{coll}_{1,2,3}(q)  \left<  \left[a I(\vec{x}) + (1-a)b    \right]  \left[a I(\vec{x'}) + (1-a)b    \right]  \left[a I(\vec{x''}) + (1-a)b    \right] I(\vec{x}''' )\right> \nonumber \\
&+& P^{coll}_{2,3,4}(q)  \left< I(\vec{x})  \left[a I(\vec{x'}) + (1-a)b    \right]  \left[a I(\vec{x''}) + (1-a)b    \right]  \left[a I(\vec{x'''}) + (1-a)b    \right] \right> \nonumber \\
&+& P^{coll}_{1,2,3,4}(q)  \left< \left[a I(\vec{x}) + (1-a)b    \right] \left[a I(\vec{x'}) + (1-a)b    \right]  \left[a I(\vec{x''}) + (1-a)b    \right]  \left[a I(\vec{x'''}) + (1-a)b    \right] \right> \nonumber, 
\end{eqnarray}
where $P^{coll}_{\o}$ is the probability that none of the four collinear points fall under the newly-deposited circle, $P^{coll}_{i,\o}$ is the probability that only the $i^{th}$ point falls under the newly-deposited circle, $P^{coll}_{i,j}$ is that probability that only the $i^{th}$ and $j^{th}$ collinear points fall under the newly-deposited circle, and so on. Because the points are collinear, it is impossible for non-neighboring pixels to fall under a given circle unless all of the pixels in between them 
also fall under that circle. Hence, there are no terms like $P^{coll}_{1,3}$ or $P^{coll}_{1,2,4}$ in the above equation, since they would require there to be ``gaps" between neighboring pixels. Alternatively, one can include those terms but note that the probabilities associated with them are zero.

To simplify Eq.(9) to the point that we can easily solve for $C^{coll}_4(q)$, we will first expand and simplify all of the average products $\left< \cdot \right>$, then compute all of the probabilities $P^{coll}_{\{ \cdot \}}$, and finally assemble all of these pieces.

\subsection{Expanding and simplifying the average pixel-value-products}

Since the circle brightnesses $b$ are zero-mean and independently drawn, each of the terms in which a single pixel is modified (the second through fifth terms in Eq. (9)) reduces to $\left<a\right> P^{coll}_{i,\o} C^{coll}_4(q)$. Similarly, expanding the terms in which 2 points fall under the circle (the sixth through eighth terms in Eq. (9)), recalling that $\left<b \right> = 0$, and performing a bit of algebra, each of those terms can be simplified to 

\begin{equation}
P^{coll}_{i,j}(q) \left[ \left< a^2 \right> C^{coll}_4(q)  + \left< (1-a)^2 \right> \left< b^2 \right> C_2(|k-m|q)   \right],
\end{equation}
 where $k \ne m$, $k,m \in \{1,2,3,4\} \backslash \{i,j\} $, $C_2(.)$ is the 2-point function that we calculated in the previous section (Eqs. (4) and (7) for power-law object size distributions), and we now denote it with a subscript 2 to avoid confusion with the 4-point function.

Assuming that the circle brightnesses are symmetrically distributed about zero (and thus $\left<b^3 \right> = 0$), the $P^{coll}_{i,j,k}$ terms in which 3 points fall under the circle reduce to

\begin{equation}
 P^{coll}_{i,j,k}(q) \left[ \left< a^3\right> C^{coll}_4(q) + \left< a(1-a)^2 \right> \left< b^2 \right> \left( C_2(q) + C_2(2q) + C_2(3q) \right) \right].
 \end{equation} 

Finally, the last term in Eq. (9), in which all 4 points fall under the new circle, simplifies to 

\begin{equation}
P^{coll}_{1,2,3,4}(q) \left[ \left< a^4\right> C^{coll}_4(q) + \left<a^2 (1-a)^2 \right> \left<b^2 \right> \left( 3C_2(q) + 2C_2(2q) + C_2(3q) \right) + \left< (1-a)^4 \right> \left<b^4 \right> \right].
\end{equation}

\subsection{Computing the probabilities $P^{coll}_{\{ \cdot \}}$}

We now require the probabilities $P^{coll}_{\o}, P^{coll}_{1,\o}, P^{coll}_{2,\o}, P^{coll}_{1,2}, P^{coll}_{2,3}, P^{coll}_{1,2,3}$, and $P^{coll}_{1,2,3,4}$. The remaining probabilities in Eq. (9) are equivalent to these because of the symmetry of the arrangement of points (and of the image ensemble).

Because all intervening pixels must lie under the circle if the bounding ones do, $P^{coll}_{1,2,3,4}(q) = P_2(3q)$, where $P_2(.)$ is the probability that 2 pixels of a given separation lie under the same circle, and is calculated in the previous section (Eq. (5) for power-law distributions of circle sizes). We will use similar arguments to obtain the other 6 probability functions that we require.

The ``triplet" probability $P^{coll}_{1,2,3}(q)$ is thus given by the probability that 3 of the (adjoining) pixels fall under the circle, minus the probability that all four pixels fall under it: $P^{coll}_{1,2,3} = P_2(2q) - P_2(3q)$. And by the same logic, $P^{coll}_{1,2} = P_2(q) - P_2(2q)$.

For the ``inner" pairs, we compute the probability of the 2 ``inner" points falling under the circle minus the probability that those two points \emph{and any adjoining ones} all fall under the circle. Thus,

\begin{eqnarray}
P^{coll}_{2,3}(q) &=& P_2(q) - P^{coll}_{1,2,3} - P^{coll}_{2,3,4} - P^{coll}_{1,2,3,4} \\
\Rightarrow P^{coll}_{2,3}(q) &=& P_2(q) - 2P_2(2q) + P_2(3q). \nonumber
\end{eqnarray}

Similarly, $P^{coll}_{1,\o}(q) = P^{\star} - P_2(q)$, where $P^{\star} = \left< s^2 \right> / L^2$ is the probability of any given point falling under the newly-deposited circle, and 

\begin{eqnarray} 
P^{coll}_{2,\o}(q) &=& P^{\star} - P^{coll}_{1,2}(q) - P^{coll}_{2,3}(q) - P^{coll}_{1,2,3}(q) - P^{coll}_{2,3,4}(q) - P^{coll}_{1,2,3,4}(q) \\
\Rightarrow P^{coll}_{2,\o}(q) &=& P^{\star}- 2P_2(q) + P_2(2q) \nonumber.
\end{eqnarray}

Finally,
 
\begin{eqnarray}
P^{coll}_{\o}(q) &=& 1 - \sum_{i} P^{coll}_{i,\o}(q) - \sum_{i,j\ne i} P^{coll}_{i,j}(q) - \sum_{i,j\ne i,k\ne i,j}P^{coll}_{i,j,k}(q) - P^{coll}_{1,2,3,4}(q) \\
\Rightarrow P^{coll}_{\o}(q) &=& 1 - 4P^{\star} + 3P_2(q) \nonumber.
\end{eqnarray}

\subsection{Assembling the pieces to find $C^{coll}_4(q)$}

Before substituting all of our results into Eq. (9) and solving for $C^{coll}_4(q)$, it will be useful to first consider the $q \gg s_0$ limit, in which we derived the 2-point function. In that limit (Eq. (6)), 

\begin{equation}
P_2(q) = A B(\alpha) \left( \frac{s_0}{L} \right)^2 \left( \frac{q}{s_0} \right)^{-(\alpha-3)} \ll 1
\end{equation}

and

\begin{equation}
C_2(q) = \frac{   B(\alpha) \left(\alpha-3 \right) \left< b^2 \right> \left< (1-a)^2  \right>}{2\left<  1-a \right>}  \left( \frac{q}{s_0} \right)^{-(\alpha - 3)} \ll 1, 
\end{equation}

so only the lowest-order terms in these quantities need to be considered. Because of the power-law nature of these functions, $C_2(2q)$ and $P_2(2q)$ have the same dependence on distance $q$ as do the $C_2(q)$ and $P_2(q)$ terms, but are smaller by a factor of $2^{-(\alpha-3)} $, and similarly for the $f(3q)$ type terms. 

Substituting all of the products and probabilities derived in the preceding subsections into Eq. (9), keeping only the lowest-order terms in $(q/s_0)^{-(\alpha-3)}$, which dominate for $q \gg s_0$, and solving for $C^{coll}_4(q)$, we find that

\begin{eqnarray}
C^{coll}_4(q) &\approx&  \frac{   B(\alpha) \left(\alpha-3 \right) \left< b^4 \right> \left< (1-a)^4  \right>}{4\left<  1-a \right>}  \left( \frac{3q}{s_0} \right)^{-(\alpha - 3)}.
\end{eqnarray}


Thus, the 4-point function for this arrangement of points (in the $q \gg s_0$ limit) has the same power-law form as does the 2-point function (Eq. 7), and it also only depends on opacity by a multiplicative pre-factor. Given that this (collinear) arrangement of points is so similar to the arrangement of points in the 2-point function (two points will always be collinear), this result is perhaps unsurprising. To test the generality of this result, we will compute the 4-point function for a square arrangement of points in the next section.

\section{Analytical calculation of the 4-point function for a square arrangement of points in the transmissive dead leaves model}

In this section, we calculate the 4-point function for our transmissive dead leaves ensemble, for the case in which the 4 points lie on the vertices of a square with edge length $q$.  Similar to the collinear arrangement of points, the symmetry in this arrangement will greatly simplify our calculations and, since it has non-trivial geometry when compared to the collinear arrangement, there is a possibility for interesting features to arise in this 4-point function that are not apparent in either the 2-point function, or the 4-point function for collinear points.

We will label these points ${1,2,3,4}$, going clockwise, and beginning in the upper left-hand corner. Similar to the calculation for the collinear case, we first list all of the possible modifications to the 4-point function, and the probabilities with which they occur. We will then simplify this expression, calculate the relevant probabilities, and use recursion to solve for the 4-point function. Similar to the previous calculations, the translation and rotation invariance of our image ensemble means that this 4-point function will depend only on the edge length of the square: $C^{square}_4(\vec{x}_1,\vec{x}_2,\vec{x}_3,\vec{x}_4) = C^{square}_4(q)$.

Enumerating all possible modifications caused by the addition of a new circle, we find that

\begin{eqnarray}
C^{square}_4(q) &=& P^{square}_{\o}(q) C^{square}_4(q)   \\
&+& 4 P^{square}_{1,\o} (q)  \left<  \left[ a I(\vec{x}_1) + (1-a)b    \right]  I(\vec{x}_2) I(\vec{x}_3) I(\vec{x}_4)  \right>  \nonumber \\
&+& 4 P^{square}_{1,2}(q)  \left<  \left[a I(\vec{x}_1) + (1-a)b    \right]\left[a I(\vec{x}_2) + (1-a)b    \right] I(\vec{x}_3)  I(\vec{x}_4)  \right> \nonumber \\
&+& 4 P^{square}_{1,2,3}(q)  \left<  \left[a I(\vec{x}_1) + (1-a)b    \right]  \left[a I(\vec{x}_2) + (1-a)b    \right]  \left[a I(\vec{x}_3) + (1-a)b    \right] I(\vec{x}_4)\right> \nonumber \\
&+& P^{square}_{1,2,3,4}(q)  \left< \left[a I(\vec{x}_1) + (1-a)b    \right] \left[a I(\vec{x}_2) + (1-a)b    \right]  \left[a I(\vec{x}_3) + (1-a)b    \right]  \left[a I(\vec{x}_4) + (1-a)b    \right] \right> \nonumber, 
\end{eqnarray}
where $P^{square}_{\o}(q)$ is the probability that none of the four corners of the square fall under the newly-deposited circle, $P^{square}_{i,\o}$ is the probability that only the $i^{th}$ corner falls under the newly-deposited circle, $P^{square}_{i,j}$ is that probability that only the $i^{th}$ and $j^{th}$ corners fall under the newly-deposited circle, and so on. The symmetries in the square configuration (all edges are equivalent, and all corners are equivalent) allow us to collapse the (equivalent) $P^{square}_{i,\o}$ terms, and similarly for the $P^{square}_{i,j}$ terms and the $P^{square}_{i,j,k}$ terms. We further note that terms like $P^{square}_{1,3}$ and $P^{square}_{2,4}$, which contain opposite corners of the circle, are omitted because it is impossible for a circle to cover diagonally opposite corners of the square without covering at least one other corner. The factors of $4$ in the above equation come in because there are 4 corners to a square, and 4 edges to a square, and ${ 4 \choose 3} = 4$ different ways to choose groupings of three of the four corners.

We can expand and simplify the averages of the products of the pixel values, as in the previous section, to find

\begin{eqnarray}
C^{square}_4(q) &=& P^{square}_{\o}(q) C^{square}_4(q)   \\
&+& 4 \left< a \right > P^{square}_{1,\o} (q)  C^{square}_4(q)   \nonumber \\
&+& 4 P^{square}_{1,2}(q) \left[ \left< a^2 \right> C^{square}_4(q)  + \left< (1-a)^2 \right> \left< b^2 \right> C_2(q)   \right] \nonumber \\
&+& 4 P^{square}_{1,2,3}(q)  \left[ \left< a^3 \right> C^{square}_4(q)  + 2 \left< a(1-a)^2 \right> \left< b^2 \right> C_2(q) + \left< a(1-a)^2 \right> \left< b^2 \right> C_2(\sqrt{2}q)   \right]     \nonumber \\
&+& P^{square}_{1,2,3,4}(q)  \left[ \left< a^4\right> C^{square}_4(q) + \left<a^2 (1-a)^2 \right> \left<b^2 \right> \left( 4 C_2(q) + 2C_2(\sqrt{2} q) \right) + \left< (1-a)^4 \right> \left<b^4 \right> \right]  \nonumber, 
\end{eqnarray}
where the function $C_2(\cdot)$ is the 2-point function we discuss in Eq. 7.

\subsection{Computing the probabilities $P^{square}_{\{ \cdot \}}$}

To finish our calculation of the 4-point function for square geometries, we require the probabilities  $P^{square}_{\o}(q)$, $P^{square}_{1,\o}(q)$, $P^{square}_{1,2}(q)$, $P^{square}_{1,2,3}(q)$, and $P^{square}_{1,2,3,4}(q)$.

For the calculation of $P^{square}_{1,2,3,4}(q)$, we first note that,  given that one of the corners of the square falls under a newly-deposited circle (with diameter $s$), the probability that all 4 points fall under it is $g_4(q/s \in [0,1/\sqrt{2}]) = \frac{4}{\pi} \left[ \cos^{-1}(q/s) - (\pi/4) + (q/s)^2 - (q/s) \sqrt{1 - (q/s)^2} \right]$.

Using the same logic (and variable substitution) as in Eq.~6, we find that

\begin{eqnarray}
P^{square}_{1,2,3,4}(q) &=& \int_{0}^{\infty} \frac{s^2}{L^2} g_4(q/s)p(s) ds \\ 
&=& A \left(\frac{s_0}{L} \right)^2 \left( \frac{q}{s_0} \right)^{-(\alpha - 3)} B_4(\alpha), \nonumber \\
\end{eqnarray}
where $B_4(\alpha) = \int_{\sqrt{2}}^{\infty} g_4(1/u) u^{2-\alpha} du$.

To derive $P^{square}_{1,2,3}(q)$, we seek the probability that 3 of the points, but not all 4, lie under the newly-deposited circle. If the two diagonal points are under the circle, so will at least one of the corners, and thus $P^{square}_{1,2,3}(q) =  (P_2(\sqrt{2} q) - P^{square}_{1,2,3,4}(q))/2$, where $P_2(x)$ is the probability that two points a distance $x$ apart lie under a newly-deposited circle, and is calculated in Eqs. 5 and 6 (above).

The ``doublet" probability $P^{square}_{1,2}(q)$ is the probability that 2, but not 3 or 4 of the points fall under the circle, and thus is given by $P^{square}_{1,2}(q) = P_2(q) - P^{square}_{1,2,3}(q) - P^{square}_{1,2,4}(q) - P^{square}_{1,2,3,4}(q) = P_2(q) - P_2(\sqrt{2}q)$.

The ``singlet" probability $P^{square}_{1,\o}(q)$ is the probability that 1, but not more, of the points fall under the circle, and is thus given by $P^{square}_{1,\o} (q)= P^{\star} - P^{square}_{1,2}(q) - P^{square}_{1,4}(q) - P^{square}_{1,2,3} (q) - P^{square}_{1,3,4}(q) - P^{square}_{1,2,4}(q) - P^{square}_{1,2,3,4}(q)$, where $P^{\star} = \left< s^2 \right> / L^2$ is the probability that a newly-deposited circle covers any given point, and is calculated in the previous sections. Simplifying this expression using our previously-derived results, we find that $P^{square}_{1,\o} (q)= P^{\star} - 2P_2(q) + \frac{1}{2} P_2(\sqrt{2} q) + \frac{1}{2} P^{square}_{1,2,3,4} (q)$.

Finally, the probability that none of the points falls under a newly-deposited circle is given by $P^{square}_{\o}(q) = 1 - \sum_i P^{square}_{i,\o}(q) - \sum_{i,j\ne i}  P^{square}_{i,j} - \sum_{i,j\ne i,k \ne i,j} P^{square}_{i,j,k} - P^{square}_{1,2,3,4} = 1 - 4 P^{\star} + 4 P_2(q) - P^{square}_{1,2,3,4}(q)$.

\subsection{Combining the pieces to find $C^{square}_4(q)$}

As in our calculation of the 4-point function for collinear points, we again consider the $q/s_0 \gg 1$ limit, in which we need only consider the lowest-order terms in $(q/s_0)^{-(\alpha-3)}$. In that limit, we find that

\begin{eqnarray}
C^{square}_4(q) &\approx&  \frac{   B_4(\alpha) \left(\alpha-3 \right) \left< b^4 \right> \left< (1-a)^4  \right>}{4\left<  1-a \right>}  \left( \frac{q}{s_0} \right)^{-(\alpha - 3)}.
\end{eqnarray}

Like the other n-point functions computed thus far, the 4-point function for square geometries is a power law with power $-(\alpha-3)$, and it depends on opacity only as a multiplicative pre-factor.  We note that, for $\alpha = 3.2$, $B(\alpha) \approx 4.014$, while $B_4(\alpha) \approx 3.581$, where these values come from numerical integration using Simpson's method. These values are similar in magnitude, and thus the 4-point function is not inherently much smaller than the 2-point function.

Finally, we note that the 2- and 4-point functions depend differently on object opacity, and thus the visible difference in the different image ensembles likely arises from the relative amplitudes of these (power-law) functions, and not any difference in their functional forms. 


\section{Numerical analysis of the transmissive fallen-leaf images}

To confirm our analytical calculations of the 2-point functions, we simulated 500-frame ensembles of $256 \times 256$ pixel images, using the procedure described in Eq.~1: circles of random size (following a power law distribution $p(s) \propto s^{-3.2}$ above the cutoff of $s_0 = 1$ pixel), brightness, and position were iteratively placed on the image frame to build up the images. For each frame, $10^6$ circles were deposited, which is the number required to cover the image surface $\sim 100$ times.

To avoid edge effects, circle centers were allowed to fall up to $256 + s/2$ pixels away from the center of the image frame, where $s$ is the circle diameter in pixels. 
We used a large maximum circle size, $s_{max} = 10^8$ pixels, because prior work~\cite{huang} 
on dead leaves models found that the functional form of the measured autocorrelation function 
approaches the analytically calculated curve only in the $ s_{\max} \to \infty $ limit. 
The heavy tail of the power-law distribution contains a non-negligible number of very large leaves, which contribute to the long-range correlations in the images. 

\begin{figure}[ht!]
\begin{center}
\includegraphics[width=3.6in]{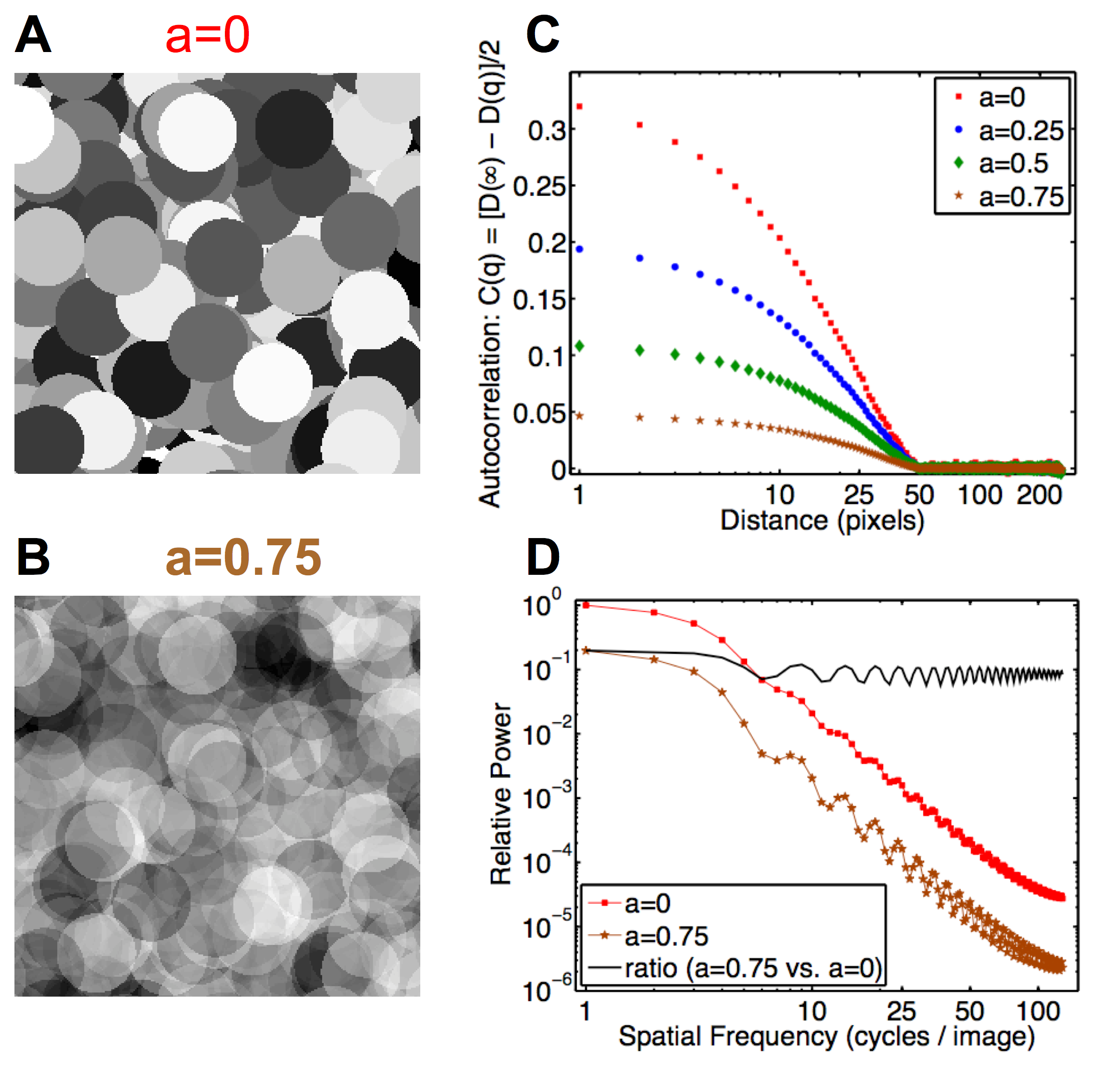}
\caption{  (Color online) \textbf{For delta-function object size distributions, opaque and transmissive dead leaves images yield different 2-point statistics.} (\textbf{A,B}) Sample images in which the leaves are all the same size ($s^{\star} = 25$ pixels), from opaque (a=0, \textbf{A}) and transmissive (a=0.75, \textbf{B}) ensembles. (\textbf{C}) The autocorrelation functions of these image ensembles do not follow power laws, and they differ from one another. (\textbf{D}) Their power spectra also differ non-trivially: the ratio between the power spectra is not constant. The ripples are at multiples of the $256/25 \approx 10$ cycles/image frequency imposed by the uniform circle size.}
\end{center}
\end{figure}

We then measured the difference functions $D(q) = \left<| I(\vec{x}) - I(\vec{x'})|^2 \right> = 2\left<I(\vec{x})^2\right> -2C(q)$ for the image ensembles. $D(q)$ is clearly related to the autocorrelation function $C(q)$, but is easier to measure~\cite{ruderman_scaling} as it is unaffected by the mean values of the individual images. We fit the measured difference functions to power law functions of the form $D(q) = \eta \times q^{\mu} + \nu$, as is suggested by our analytical calculations (Eq.~7). The best-fit parameters $(\eta,\mu,\nu)$ for the image ensembles with $a= \{0,0.25,0.5,0.75 \}$ were $(-0.48 \pm 0.01, -0.24 \pm 0.04, 0.69 \pm 0.03)$, $(-0.32 \pm 0.01, -0.23 \pm 0.03, 0.41 \pm 0.02)$, $(-0.191 \pm 0.004, -0.22 \pm 0.03, 0.23 \pm 0.01)$, $(-0.086 \pm 0.002, -0.21 \pm 0.02, 0.098 \pm 0.005)$, respectively, where the uncertainties represent $95\%$ confidence intervals. These values are in good agreement with the analytical calculations that predict $\mu = -0.2$ for all ensembles, and $\nu = \{ 0.66,0.396,0.22,0.094 \} $ for the ensembles with $a= \{0,0.25,0.5,0.75 \}$, respectively. The correlation functions shown (Fig.~2D) are the measured difference functions subtracted from the constants $\nu$ measured in the fit: $C(q) = [\nu - D(q)]/2$. These correlation functions are power-law functions of distance (linear on the log-log plot), and differ by a multiplicative constant. Similarly, the power spectra of the image ensembles (Fig.~2E), differ only by a multiplicative constant for low spatial frequencies, where the $q \gg s_0$ approximation holds.

Fig.~3 demonstrates that the 2-point function is affected substantially by leaf opacity for delta-function size distributions. In particular, the modulation depth of the ``ripples" in the power spectra depend on the leaf opacity, and thus the opacity does not modify the power spectra simply by a multiplicative factor. The procedures used to generate the data shown in Fig.~3 were the same as for the power-law object size distribution, except for the different distribution of object sizes.

\section{A more realistic model of radiological images}

\begin{figure}[h!]
\begin{center}
\includegraphics[width=3.6in]{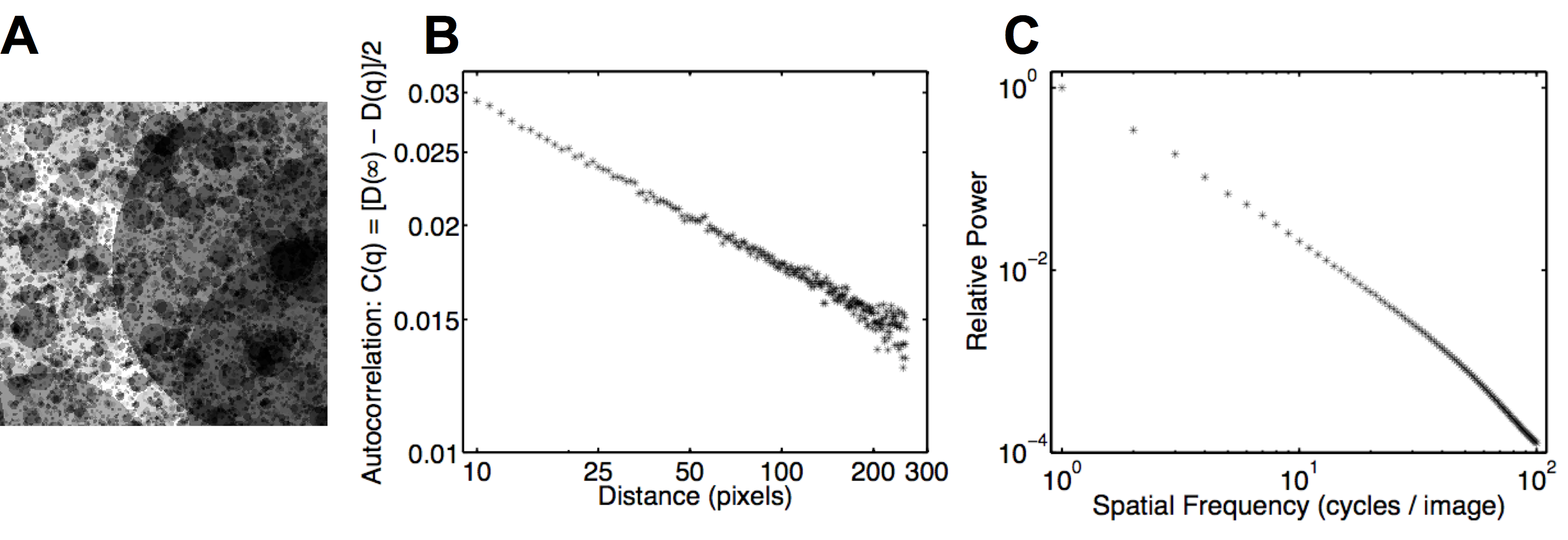}
\caption{ \textbf{A shadowing dead leaves model with finite optical depth also exhibits scale invariant 2-point statistics for power law object size distributions.} (\textbf{A}) An example image from a dead leaves model with the same power law distributed leaf sizes as in Fig.~2, in which each leaf leaves a shadow by multiplying the brightness of the pixels it subtends by a factor no greater than one, drawn uniformly within $[0.5,1]$. Unlike the previous models, each pixel starts out at full brightness, and only a finite number of circles is added to generate the image. The autocorrelation function (\textbf{B}) and power spectrum (\textbf{C}) of this ensemble show scale invariance (for relatively low frequencies, which corresponds to $q \gg s_0$), just like the previous models.}
\end{center}
\end{figure}

Our transmissive dead leaves model is not a perfect model for radiological images.  Image formation in mammograms and other projectional radiographs results from the partial blockage of a roughly uniform illumination of x-rays due to local regions of dense tissue, unlike our dead leaves model. 
Moreover, imaged tissue is typically much thinner than the path length required to fully block the x-rays throughout the image, unlike the effectively infinite optical depth of our ``additive" transparent dead leaves model (Eq. (1)). 

It is thus natural to ask whether our conclusions about variable object opacity generalize to these types of images. Analytically computing the 2-point statistics for this radiographic model is more involved than for the infinite depth models, since recursion is more complex in this case. For this reason, we chose to verify via simulation that the qualitative results from our analytical calculations hold for these types of images.

Fig.~4A shows a typical image from a shadowing dead leaves model with finite optical depth and the same power law leaf size distribution as in the previous models.  To generate these model images, a uniform background illumination (of 1) was imposed across the whole image. Randomly sized and located circles were then deposited onto the image plane, with each leaf multiplying the brightness of the pixels it subtends by a factor drawn uniformly within $[0.5,1]$. The circle sizes were drawn from the same power law distribution as in the previous simulations, and the simulation code was thus very similar. 

For an ensemble of these ``radiographic" images, the empirically measured 2-point function (Fig.~4B) and power spectrum (Fig.~4C) exhibit the same power laws as we found for our previous models (Figs.~2,3), suggesting that our calculation holds more generally than for the specific model for which we performed the analytical calculations.

Intuitively, one might expect the same scale-invariant 2-point function for this model as for the previous one since no new length scale has been introduced. 

\section{Conclusions}

For the special case of power-law object size distributions, object opacity does not affect the form of either the 2- or 4-point functions, or the power spectrum of images: it is manifest only by a multiplicative constant in these power-law functions. Ours is the first analytic calculation that demonstrates these facts, and thus deepens our understanding of image statistics.

For object size distributions other than power-law, object opacity can (potentially dramatically) alter the low-level image statistics. Occlusion is important for natural image formation, but we find that it does not change the form of the power spectrum. Since images formed by opaque leaves that are all the same size have oscillatory, non-power-law, power spectra (Fig.~3), and transmissive leaves can yield power law power spectra (Figs.~2 and 4), occlusion is likely not responsible for scale invariance of images. We propose that the universality of power law power spectra in both occlusive imaging environments, such as natural photographic images, and transmissive ones, such as mammography, is likely due to power-law object size distributions in both settings. 


\acknowledgments
JZ's contribution to this work was supported by an international student research fellowship from the Howard Hughes Medical Institute (HHMI). This material is based upon work supported by a National Science Foundation Graduate Research Fellowship to DP under Grant No. DGE Ð 11-44155. MRD thanks the Hellman Family Foundation, the James S. McDonnell Foundation, and the McKnight Foundation for support.

\end{document}